# Intrinsic Nanoscale Inhomogeneities in Transition Metal Oxides

J.W. Lynn[1], D. N Argyriou[2], Y. Ren[3], Y. Chen[1,4], Y. M. Mukovskii[5], and D. A. Shulyatev[5]

[1]NIST Center for Neutron Research,
National Institute of Standards and Technology, Gaithersburg, MD 20899

[2]Hahn-Meitner-Institut, Glienicker Strasse 100, Berlin D-14109, Germany

[3]X-ray Science Division, Argonne National Laboratory, Argonne, IL 60439

[4]Department of Materials Science and Engineering, University of Maryland, College Park, MD 20742

[5]Moscow State Steel and Alloys Institute, Moscow 119991, Russia

Abstract

Neutron elastic, inelastic and high energy x-ray scattering techniques are used to explore the nature of the polaron order and dynamics in $La_{0.7}Ca_{0.3}MnO_3$. Static polaron correlations develop within a narrow temperature regime as the Curie temperature is approached from low temperatures, with a nanoscale correlation length that is only weakly temperature dependent. The static nature of these short-range polaron correlations indicates the presence of a glass-like state, similar to the observations for the bilayer manganite in the metallic-ferromagnetic doping region. In addition to this elastic component, inelastic scattering measurements reveal dynamic correlations with a comparable correlation length, and with an energy distribution that is quasielastic. The elastic component disappears at a higher temperature T*, above which the correlations are purely dynamic. The overall behavior bears a remarkable resemblance to the magnetic fluctuation spectrum recently observed in underdoped cuprates.

## I. Introduction

Oxides containing perovskite building blocks, such as doped $LaMnO_3$ that exhibits colossal magnetoresistivity [1], relaxor ferroelectrics [2], or the superconducting cuprates [3], are highly correlated electron systems that exhibit a wide variety of interesting physical properties. One of the emerging features of these materials is their propensity to form intrinsically inhomogeneous structures on a variety of length scales, from full scale phase separation to nanoscale polarons in the manganites, polar nanoregions in relaxors, and spin and charge stripes in cuprates. The $La_{1-x}Ca_xMnO_3$ (LCMO) material of particular interest here is a prototype system that is a metallic ferromagnet in the regime $0.2<x<0.5$ [4-7]. Colossal magnetoresistivity (CMR) is observed in this regime, and is associated with the formation of nanoscale polarons that develop at elevated temperatures, truncating the ferromagnetic metallic state and driving the transition first order [8-24].

In the region around $x\sim0.3$, polarons take the form of correlations with an ordering wave vector of $\sim(1/4,1/4,0)$. Indeed this type of ordering becomes long range at half doping, where equal numbers of $Mn^{3+}$ and $Mn^{4+}$ form a charge and orbitally ordered structure know as CE-type [4,5]. The ordering takes the form of alternating $Mn^{3+}$ and $Mn^{4+}$ ions where the magnetic coupling between filled $Mn^{3+}$ $3d_{3z^2-r^2}$ and empty $Mn^{4+}$ $e_g$ states forms ferromagnetic zig-zag chains, while the coupling between chains is antiferromagnetic. For $x<1/2$ this structure is frustrated and it is observed in the insulating state of manganites in the form of nanoscale correlations [12,13]. While these CE correlations characterize the insulating state just above $T_C$, the onset of ferromagnetism leads to their melting and no diffuse scattering is observed far below $T_C$ [8,12,13, 20].

We have been investigating the structure and dynamics of this ferromagnetic-metallic to paramagnetic-insulating transition using neutron and x-ray scattering techniques. As the paramagnetic state is entered, a purely elastic component to the structural polaron scattering signals the development of the correlated polaron glass phase [12,13,20], in a manner similar to that observed in the bilayer manganite system [16-19]. The polaron correlation length in this phase is around a nanometer and is only weakly temperature dependent, while the strength of the elastic scattering diminishes with increasing temperature until the static polarons disappear at a higher temperature T*. The correlations remain above this temperature, but are then purely dynamic in character. The statics and dynamics of this scattering bear a remarkable similarity to the magnetic fluctuation spectrum recently observed in underdoped cuprates, suggesting that the underlying behavior has a similar origin [25].

## II. Experimental Details

A 1.5 g single crystal of $La_{0.7}Ca_{0.3}MnO_3$ was grown by the floating zone technique [26] and was used for the triple axis neutron measurements on the BT-2, BT-7, and BT-9 spectrometers at the NIST Center for Neutron Research, employing pyrolytic graphite monochromator, analyzer, and filter at a fixed energy of 14.7 or 14.8 meV. A small piece of the original crystal [20] was used for the x-ray measurements at the 11-ID-C beam line at the Advanced Photon Source, with a high x-ray energy of 115 keV. The crystals were mounted in a closed cycle refrigerator for temperature control. This is the highest composition where single crystals have been successfully grown, and the measured Curie temperature $T_C$ is 257 K. At this composition the crystal structure is orthorhombicc)
(*Pnma*) with $a$=5.4654 Å, $b$= 5.4798 Å, and $c$= 7.7231Å. The data were taken in the (*h,k,0*) scattering plane in this notation, where the polaron correlation peaks are located at half-integer positions like (½,0,0). The distortion away from the ideal cubic perovskite structure is small (but important) and the domains are equally populated, and hence the cubic frame of reference is often employed where we have $a\approx3.87$ Å and the nearest-neighbor manganese atoms are along the [100]-type directions. In this notation the polaron correlation peaks occur at (¼,¼,0) and equivalent positions. At higher (x≥½) doping these peaks are genuine Bragg peaks indicative of CE-type [4,5] long range order, and thus these polaron peaks are often denoted CE-type peaks.

## III. Results

### A. Static Polaron Correlations

An example of the observed diffuse x-ray scattering data is shown in Fig. 1. In addition to the fundamental Bragg peak observed at the integer position, we observe a broad peak at the half-integer position (orthorhombic notation) (Fig. 1(a) and (c)). This scattering is strongly temperature dependent. This is shown in Fig. 1c and d where we have constructed a map of reciprocal space using line scans over this (H,K,0) region. While the map at 300 K (Fig. 1(c)) shows the total scattering at this temperature (see below), in Fig 1(d) we show the difference plot of the scattering, where we have taken "background" below $T_C$, at 220 K, and subtracted it from the scattering observed at 300 K. This difference plot demonstrates that there are two new components to the scattering, the peak at the half-integer position and a diffuse scattering component that peaks at the position of the fundamental Bragg peak. Fig. 1(e) and (f) show these two components in detail. Here we plot radial scans away from the fundamental peak, and scans directly across the peak at the half-integer position as a function of temperature in the form of a ($q$,T) map. These data and in particular the data in Fig. 1(f) at 220 K show that there is a quite strong ridge of scattering at this temperature, while the data in the radial scan show only a broad shoulder (at most) below the ferromagnetic transition. This peak rapidly develops in the vicinity of the ferromagnetic transition on warming and has a maximum at or very near to $T_C$ as shown by the scattering at 253 K (Fig. 1(a) and (b)). The scattering then decreases with further increase of temperature.

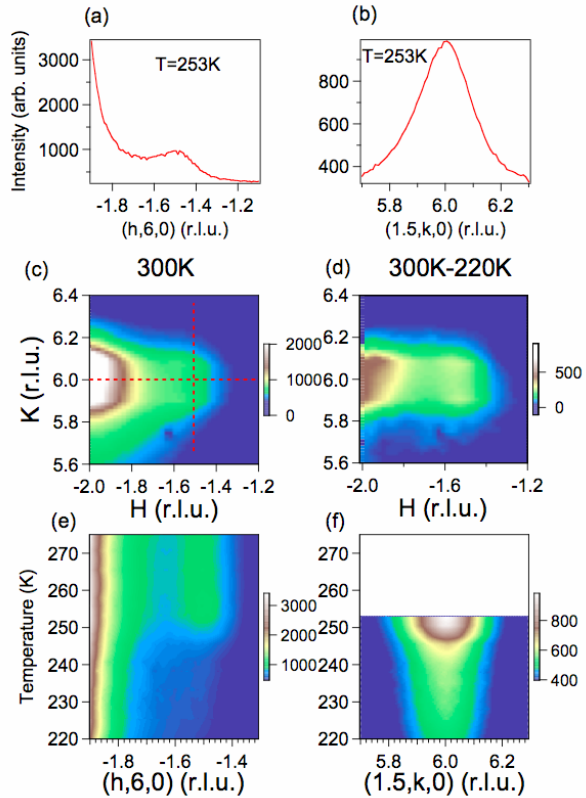

**Figure 1.** (color online) Diffuse scattering observed in a single crystal of $La_{0.7}Ca_{0.3}MnO_3$ using high energy synchrotron x-rays measured on beam line 11ID-C at APS. The data are cuts through the observed scattering centered for the (0,1/2,0) peak (in orthorhombic notation; in cubic notation this is the (1/4,1/4,0) "CE" position). Panel (a) shows radial cuts from the fundamental Bragg peak, while panel (b) shows a transverse scan through the CE peak. These are example scans measured at 253 K. In panel (c) we show an intensity map around the (6,-2,0) fundamental Bragg peak (orthorhombic notation), and we observe an additional broad peak at the half-integer position in the transverse direction (6,-1.5,0). There is no significant diffuse scattering at the (6.5,-2,0) position, as expected for a CE-type lattice distortion. In panel (d) we show a subtraction of the scattering below the transition (220K) from the scattering observed above (300 K). In panels (e) and (f) we show the temperature dependence of the radial and transverse scans in the form of a (*q*,T) map. These maps were made by repeating the measurements shown in panels (a) and (b), while varying temperature. It is evident from these data that the polaronic scattering rapidly increases on warming through $T_C$, where a diffuse peak appears at (1/4,1/4,0).

X-ray scattering has no energy resolution in these measurements, and thus integrates the scattering over all energies to obtain the instantaneous correlation function. The acoustic phonons around the fundamental Bragg peaks give a temperature-dependent contribution to the diffuse scattering that is proportional to $1/q^2$, and we used this form along with a Lorentzian peaked at H~-1.5 to quantitatively analyze these data. This form for the diffuse scattering provides a good fit to the data, and the results for the fits are shown in Fig. 2. The intensity of the polaron peak at H~-1.5 (Fig. 2a) is seen to sharply peak at $T_C$, and then is found to decrease with further increase of T, while the width of the scattering above $T_C$ (Fig. 2b) is only weakly dependent on T. Indeed at the higher temperatures, when the strength of this scattering becomes weak, we found that it was necessary in the x-ray data to fix the width to the highest temperature value that we could refine ($\Delta h$=0.105 r.l.u.). This overall behavior is in good agreement with previous neutron data for this polaron scattering, where fixing the width at higher temperatures was not necessary [12-14]. We remark that the similarity of the x-ray and neutron measurements, as well as neutron polarized beam data, demonstrate that this scattering originates from the lattice as opposed to magnetic scattering [12-14, 16].

Finally, we note that the fitted width becomes substantially broader below $T_C$, but there is still scattering observed, and even though the overall intensity is drastically reduced the integrated intensity is still significant. It would be interesting to investigate this component as a function of doping to determine if it is absent at optimal doping, but for LCMO single crystals are not available in this doping regime.

The intensity of the diffuse scattering is also shown in the figure, and we see that there is an abrupt increase in this component of the scattering at the ferromagnetic transition temperature. Thus this component of the scattering has two contributions, one from the usual temperature dependence of the acoustic phonons (thermal diffuse scattering (TDS)), and another (Huang) component originating from lattice relaxations around uncorrelated "single polaron" and correlated polarons. We remark that away from the half-integer polaron peaks, we do not observe any anomalies as a function of temperature in the neutron inelastic scattering measurements, and thus the change in this x-ray scattering component at $T_C$ cannot be attributed in any way to a change in the (acoustic) phonon dispersion. Similar conclusions have been obtained for the case of the bilayer manganite system [16, 18]. For the neutron measurements the elastic and inelastic contributions can be resolved, and the different temperature dependence to the single and correlation polaron scattering has been established [14]. The solid squares in Fig. 2(a) indicate the elastic polaron intensity measured with neutron energy analysis, scaled to the x-ray intensity at the maximum. The close correspondence of the x-ray and neutron intensities as a function of temperature demonstrates that the x-ray scattering is primarily elastic in origin.

There are two interesting features to these two components of the diffuse scattering that originate from the polarons. The first is that for the polaron correlation peak, the width of the scattering is quite broad, corresponding to a real space correlation length of ~1 nm (at its maximum value at $T_C$). Moreover, the width is also only weakly temperature dependent [12-15, 20, 21], showing a relatively small anomaly around $T_C$. This behavior differs qualitatively from what would be expected in a classic second-order phase transition, where one would expect the width to rapidly decrease as the transition is approached from above, indicating that the correlations are increasing in size. The correlation length would then diverge (peak would become resolution-limited experimentally) at the phase transition, marking the onset of long range order. The observed weak dependence with temperature indicates that the phase transition is first-order in nature, with two-phase coexistence, and this first-order behavior has been investigated in depth [8-15, 20-24]. We remark that the magnetic correlation length is also ~1 nm and weakly temperature dependent [8, 12-13, 20, 27].

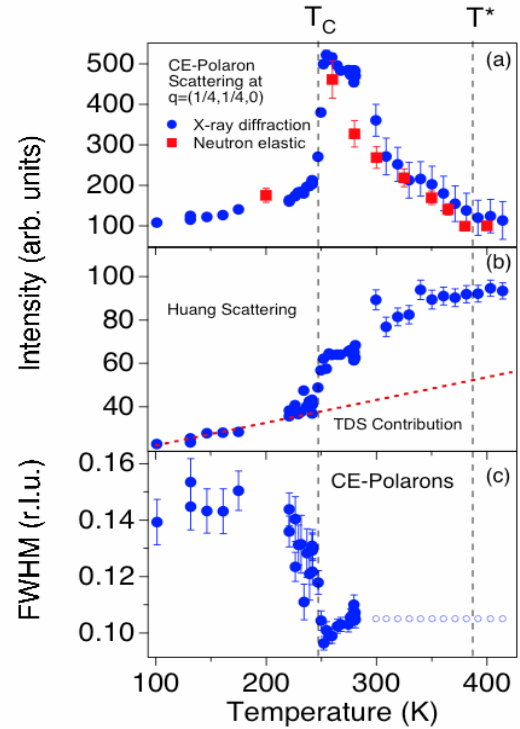

**Figure 2.** (color online) (a) Integrated intensity (filled circles) vs. T for the CE-peak, obtained from fits to the x-ray scattering. The elastic component of the correlated polaron scattering measured with energy-resolved neutrons (filled squares) closely follows the x-ray intensity and disappears at the same T, defining a T* above which all the correlations are dynamic, and below which there is a frozen component that defines a polaron glass. Both types of scattering show a jump in the scattering at the ferromagnetic-metallic to paramagnetic-insulating transition. (b) The Huang scattering determined from $1/q^2$ fits to the data is shown as a function of temperature. The scattering contains a linear term in T that arises from thermal diffuse scattering (dashed red line). At sufficiently high T the Huang scattering becomes approximately equal to that of the CE-polarons, indicating that the x-ray intensity in this temperature range is purely from uncorrelated single polarons. (c) Full width at half maximum in r.l.u. as a function of T, for the CE-peak obtaining from Gaussian fits to the x-ray diffuse scattering. For T>285K the CE-peak was observable in the data but the width could not be refined independently. For the purpose of fitting it was fixed to a value of $\Delta h$=0.105 (unfilled circles).

The second feature concerns the comparison between the Huang and "correlated polaron" scattering. At sufficiently high temperatures we see that these two cross sections become approximately equal, indicating that the wave-vector-dependent peak in the elastic scattering has vanished. This indicates that at sufficiently high temperatures polaronic correlations have broken down, while the presence of Huang scattering is indicative of single polarons. We use this feature to define a T*≈380 K, where for T>T* there are no longer any static polaron correlations evident. This indicates that the glass phase has melted, as has been found in the bilayer manganite system [18]. As we will see below, there are still dynamic correlations present above T*, as is the case for the bilayer system, but the static order is gone.

## B. Dynamic Polaron Correlations

We now turn to the dynamics of the scattering as revealed by inelastic neutron scattering measurements. The energy dependence of the scattering at low energies is shown in Fig. 3 for a wave vector at the peak position (($\frac{1}{4}$,$\frac{1}{4}$,0) position in the cubic notation used here) of the polaron scattering. To obtain these data we have subtracted the scattering observed at 200 K, where the elastic component of the polaron scattering is small. At this temperature the scattering at the elastic position is then dominated by the purely elastic, nuclear incoherent scattering. This incoherent nuclear cross section contains the Debye-Waller factor and so has some temperature dependence, and thus to make this subtraction we have reduced the level of this "background" scattering by the Debye-Waller factor obtained from powder diffraction [28]. We remark that this is a significant correction at these elevated temperatures compared with the bilayer material, as noted in that work [18] and in earlier results [16].

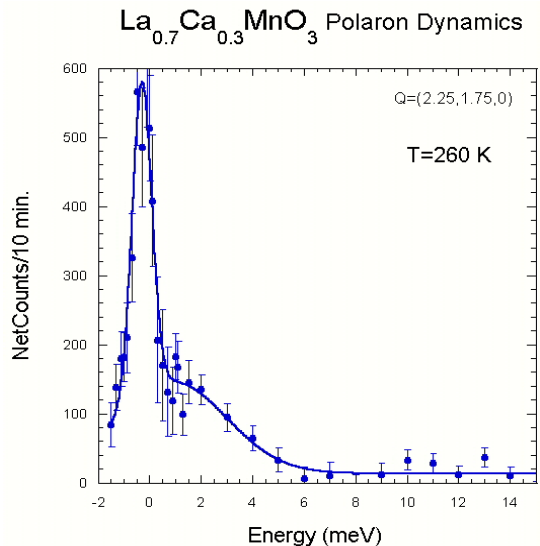

**Figure 3.** (color online) Energy dependence of the scattering at the wave vector position of the polaron scattering. There are two components of the scattering evident, a purely elastic component, and inelastic scattering that peaks at E=0 (quasielastic scattering).

There are two components of the polaron scattering evident in the energy dependence of the scattering as shown in Fig. 3. One is a purely elastic component, that is, a component that is resolution limited and thus corresponds to static order [29]. Recalling that the elastic scattering is quite broad in wave vector, the interpretation of the behavior of the system is that a static, polaron glass phase has formed, where the range of the order is only ~1 nm. The second component of scattering is dynamic in origin, but also peaks at E=0. This is then a quasielastic component of the scattering, and has an energy width of $\Gamma \approx 3$ meV at this temperature. Quasielastic scattering corresponds to a finite lifetime of the CE-correlations. On cooling the liefetime of these correlations increases leading up to a the glass phase below T*. We interpret this increase to be proportional to the hoping rate of the charge carriers which diffuse through the lattice.

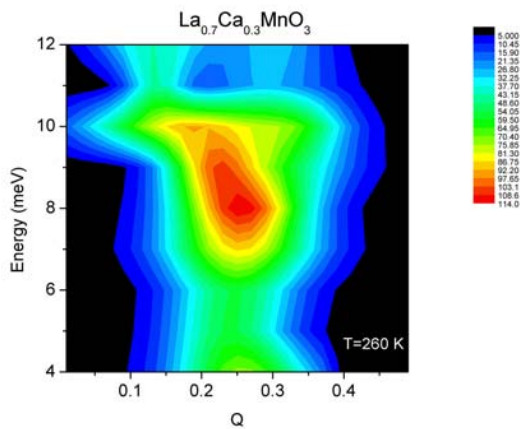

**Figure 4.** (color online) Wave-vector—Energy map of the CE-type inelastic scattering at a temperature of 260 K, just above the Curie temperature. The quasielastic scattering at low energies peaks at higher energies, and appears to be associated with a soft, broadened phonon. These data were taken with a fixed incident energy of 14.7 meV in energy gain. Low temperature (5 K) background have been subtracted, and the intensities have been corrected for the volume of the resolution function.

A map of the intensity of the inelastic scattering at higher energies in the region of these CE-type peaks as a function of **Q** and E is shown in Fig. 4 at a temperature just above $T_C$, where the scattering is strongest. We see that the dynamical scattering extends to quite high energies. There is also a peak in the scattering around 8 meV, which originates from a phonon mode that is also observed below $T_C$. Figure 5 shows constant-E scans through this high energy scattering, and indicates that the polaron scattering at lower energies evolves into single stronger-intensity peak, while at higher energies we see two peaks that disperse. Although the temperature for these measurements are relatively high and therefore the interpretation of these scans is somewhat difficult at this stage, we speculate that they may be related to optic phonon anomalies associated with polarons [30]. We remark that this ~8 meV phonon is observed to be sharp in the ferromagnetic-metallic state, with only a modest shift to higher energies. A similar interaction with the transverse acoustic phonon has been observed $La_{0.75}(Ca_{0.45}Sr_{0.55})_{0.25}MnO_3$ [21], and soft/heavily-damped phonon behavior was also found for the bilayer manganite but the phonon energy was much higher and the relationship between the phonon mode and the polaron scattering was not as clear, and the transverse acoustic phonon was not significantly affected [18]. However, because of the elevated temperatures involved in these LCMO measurements (compared to the bilayer system), the Debye-Waller corrections noted above, and the strong temperature-dependent damping of the phonon, we found that a detailed analysis of the temperature dependence of the energy width of the quasielastic scattering was not possible.

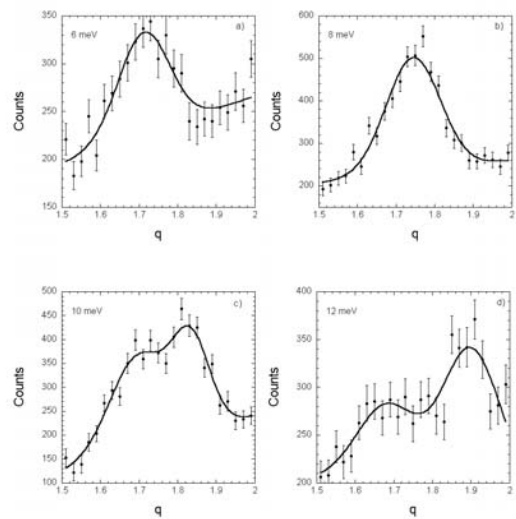

**Figure 5.** Transverse (as in Fig. 2b) constant-E scans through the CE-type scattering at 260 K for four different energies. A single peak is observed at lower energies and appears to be associated with a soft phonon, while at higher energies two dispersive peaks are observed.

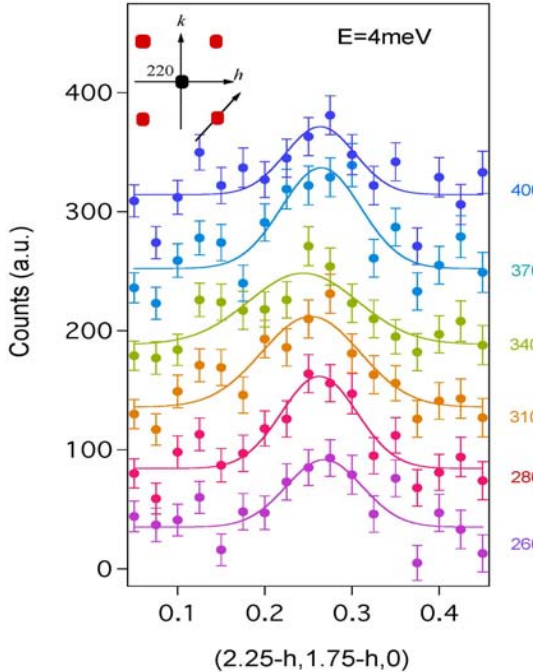

**Figure 6.** (color online) Temperature dependence of the inelastic scattering at an energy transfer of 4 meV. This scattering abruptly appears on warming near the ferromagnetic-metallic to paramagnetic-insulating phase transition. The wave vector width of these dynamic correlations is essentially temperature independent up to the highest temperature measured, indicating that strong dynamic correlations persist to high temperature, even though the polaron glass melts (defined when the purely elastic component has vanished).

In Fig. 6 we show the wave vector dependence of the lower-energy inelastic scattering at an energy transfer of 4 meV, at a series of temperatures. Well into the ferromagnetic state we observe a flat distribution of "background" scattering (not shown) due to multiphonon and multimagnon scattering, while above $T_C$ we see a relatively broad but well defined peak that abruptly appears on warming near the ferromagnetic-metallic to paramagnetic-insulating phase transition. The wave vector width of the scattering corresponds to a length scale in real space of ~1 nm, very similar to the static portion of the scattering found in this temperature regime [12-14, 20]. Interestingly, the width of these dynamic correlations is again only weakly dependent on temperature up to the highest temperature measured, indicating that strong dynamic correlations persist to high temperature, even though the polaron glass has melted (defined when the purely elastic component has vanished).

## IV. Discussion

The development of the polaron ordering in the paramagnetic state has been studied in some detail in the "cubic" (La-CaMnO$_3$-type) systems previously [12-14, 20, 21, 31]. This scattering abruptly appears in the paramagnetic state, and is directly related to the colossal magnetoresistive behavior in that the polaron intensity closely tracks the resistivity as a function of applied magnetic field [12-14]. In particular it has been found both experimentally [21] and theoretically [32] that there is no CMR effect when only single polarons are present. The polaron dynamics have also been investigated in the bilayer manganite. Here the two-dimensional nature of the spin dynamics [33] reduces the Curie temperature, allowing measurements at significantly lower temperature. The bilayer system also has the advantage of tetragonal symmetry, rather than the present orthorhombic system which is twinned. In the bilayer material a polaron glass phase, identified by a purely elastic component of the polaron scattering, was clearly established [18]. The present study establishes that the same basic polaron glass behavior is found in the (La-Ca)MnO$_3$ "cubic" class of materials. The glass melts at higher temperature, but dynamic correlations persist.

In LCMO the sudden development of these static CE-type polaron structures has important implications for the nature of the phase transition. In particular, when these static polarons form they have the effect of trapping electrons and driving the system into the insulating state, which is accompanied by the truncation of the ferromagnetic phase in a discontinuous manner [8,10,12-15, 20,22]. Thus the transition has a strong first order component [8-15,20-24, 34-37]. At higher doping the transition becomes second order [23], while at small doping the glass transition is thought to evolve into the Jahn-Teller driven cubic to orthorhombic transition for $LaMnO_3$ [38], where the transition is long range in nature. In Figure 7 we place the dynamic and glassy polaronic region on the (x,T) phase diagram for $La_{1-x}Ca_xMnO_3$ using data from this work and that in references [6-8, 12-14, 20,23, 26, 38-39]. We suggest that this polaron glass is an additional phase that occurs in the topology of the phase diagram. We remark that in the case of $La_{1-x}Sr_xMnO_3$ and $La_{1-x}Ba_xMnO_3$, (1/4,1/4,0)-type static polaron structures are not evident. In this case the ferromagnetic to paramagnetic transition is found to be a conventional $2^{nd}$ order phase transition, where (for example) conventional scaling behavior for the magnetization is observed [40]. However, there may still be dynamic polaron correlations in these materials, and that will be the subject of separate studies.

In the polaron glass phase of LCMO we have two components to the low energy scattering, the elastic scattering due to the static order, and the quasielastic scattering due to polaron diffusion. These two low-energy components of scattering bear a remarkable resemblance to the magnetic fluctuation spectrum recently observed in underdoped $YBa_2Cu_3O_{6.35}$ in the vicinity of the antiferromagnetic (π,π)

position [25]. Indeed the data for the

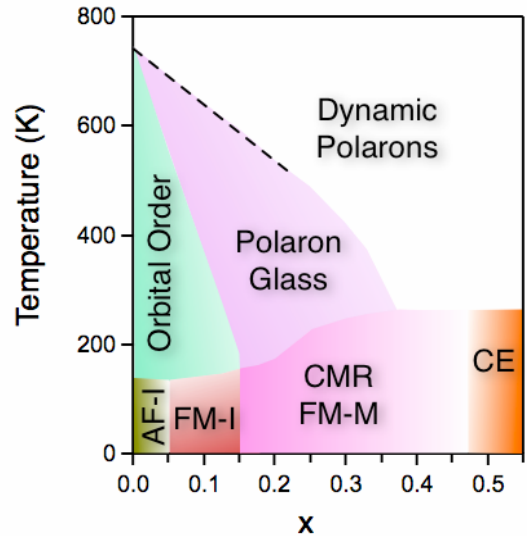

**Figure 7**. (color online) Phase diagram for $La_{1-x}Ca_xMnO_3$, were we have sketched the proposed polaronic glass phase and how it evolves into the long-range ordered Jahn-Teller transition for the undoped compound. Basic ferromagnetic-metallic to paramagnetic-insulating, ferromagnetic-insulating to paramagnetic insulating, and antiferromagnetic to paramagnetic phase transitions are after [4, 6, 7], noting that at lower x the transitions are also dependent on the oxygen concentration and hence the heat treatment of the samples [28]. The ferromagnetic-paramagnetic transitions are from our work and references [8, 12, 14, 20, 28, 39], while above optimal doping the transition becomes second order [23]. The polaron glass phase is based on the present work.

lattice scattering we observe in Fig. 3 looks identical to Fig. 1a of that work, indicating that transition metal oxides with the basic perovskite building blocks exhibit important commonalities of the underlying physical mechanisms that control the properties of these highly correlated electron systems.

## V. Acknowledgments

We would like to thank Elbio Dagotto, Pengcheng Dai, Jaime Fernandez-Baca and Daniel Khomski for helpful conversations. Use of the Advanced Photon Source was supported by the U.S. Department of Energy, Office of Science, Office of Basic Energy Science, under Contract No. W-31-109-ENG-38.

**References**

[1] E. Dagotto, Science **309**, 257 (2005); Y. Tokura and N. Nagaosa, Science **288**, 462 (2000).
[2] G. Burns and F. H. Dacol, Phys. Rev. B**28**, 2527 (1983).
[3] S. A. Kivelson, I. P. Bindloss, E. Fradkin, V. Oganesyan, J. M. Tranquada, A. Kapitulnik and C. Howald, Rev. Mod. Phys. **75**, 1201 (2003).
[4] E. O. Wollan and W. C. Koehler, Phys. Rev. **100**, 545 (1955).
[5] J. B. Goodenough, Phys. Rev. **100**, 564 (1955).
[6] P. Schiffer, A. P. Ramirez, W. Bao, and S.-W. Cheong, Phys. Rev. Lett. **75**, 3336 (1995).
[7] S.-W. Cheong and C. H. Chen, in *Colosssal Magnetoresistance,Charge Ordering, and Related Properties of Manganese Oxides*, edited by B. Raveau and C. N. R. Rao (World Scientific, Singapore, 1998), p. 241.
[8] J. W. Lynn, R. W. Erwin, J. A. Borchers, Q. Huang, A. Santoro, J.-L. Peng, and Z. Y. Li, Phys. Rev. Lett. **76**, 4046 (1996). J. W. Lynn, R. W. Erwin, J. A. Borchers, A. Santoro, Q. Huang, J.-L. Peng, and R. L. Greene, J. Appl. Phys. **81**, 5488 (1997).
[9] W. Archibald, J.-S. Zhou, and J. B. Goodenough, Phys. Rev. B**53**, 14445 (1996).
[10] J. A. Fernandez-Baca, P. Dai, H. Y. Hwang, C. Kloc, and S-W. Cheong, Phys. Rev. Lett. **80**, 4012 (1998).
[11] J. Mira, J. Rivas, F. Rivadulla, C. Vazquez-Vazquez, and M. A. Lopez-Quintela, Phys. Rev. B**60**, 2998 (1999).
[12] C. P. Adams, J. W. Lynn, Y. M. Mukovskii, A. A. Arsenov, and D. A. Shulyatev, Phys. Rev. Lett. **85**, 3954 (2000).
[13] P. Dai, J. A. Fernandez-Baca, N. Wakabayashi, E. W. Plummer, Y. Tomioka, and Y. Tokura, Phys. Rev. Lett. **85**, 2553 (2000).
[14] J. W. Lynn, C. P. Adams, Y. M. Mukovskii, A. A. Arsenov, and D. A. Shulyatev, J. Appl. Phys. **89**, 6846 (2001).
[15] C. S. Nelson, M. v. Zimmermann, Y. J. Kim, J. P. Hill, D. Gibbs, V. Kiryukhin, T. Y. Koo, S.-W. Cheong, D. Casa, B. Keimer, Y. Tomioka, Y. Tokura, T. Gog, and C. T. Venkataraman, Phys. Rev. B**64**, 174405 (2001).
[16] L. Vasiliu-Doloc, S. Rosenkranz, R. Osborn, S. K. Sinha, J.W. Lynn, J. Mesot, O. H. Seeck, G. Preosti, A. J. Fedro, and J. F. Mitchell, Phys. Rev. Lett. **83**, 4393 (1999); L. Vasiliu-Doloc, R. Osborn, S. Rosenkranz, J. Mesot, J. F. Mitchell, S. K. Sinha, O. H. Seeck, J. W. Lynn, and Z. Islam, J. Appl. Phys. **89**, 6840 (2001).
[17] S. Shimomura, N. Wakabayashi, H. Kuwahara, and Y. Tokura, Phys. Rev. Lett. **83**, 4389 (1999).
[18] D. N. Argyriou, J. W. Lynn, R. Osborn, B. Campbell, J. F. Mitchell, U. Ruett, H. N. Bordallo, A. Wildes, and C. D. Ling, Phys. Rev. Lett. **89**, 036401 (2002).
[19] B. J. Campbell, S. K. Sinha, R. Osborn, S. Rosenkranz, J. F. Mitchell, D. N. Argyriou, L. Vasiliu-Doloc, O. H. Seeck, and J. W. Lynn, Phys. Rev. B**67**, 020409(R) (2003).
[20] C. P. Adams, J. W. Lynn, V. N. Smolyaninova, A. Biswas, R. L. Greene, W. Ratcliff, II, S-W. Cheong, Y. M. Mukovskii, and D. A. Shulyatev, Phys. Rev. B**70**, 134414 (2004).
[21] V. Kiryukhin, A. Borissov, J. S. Ahn, Q. Huang, J. W. Lynn, and S-W. Cheong, Phys. Rev. B**70**, 214424 (2004).
[22] T. J. Sato, J. W. Lynn, and B. Dabrowski, Phys. Rev. Lett. **93**, 267204 (2004).
[23] D. Kim, B. Revaz, B. L. Zink, and F. Hellman, J. J. Rhyne, and J. F. Mitchell, Phys. Rev. Lett. **89**, 227202 (2002).
[24] K. H. Kim, M. Uehara, and S-W. Cheong, Phys. Rev. B**62**, R11945 (2000); T. J. Sato, J. W. Lynn, Y.-S. Hor, and S-W. Cheong, Phys. Rev. B**68**, 214411 (2003). L. Downward, F. Bridges, S. Bushart, J. J. Neumeier, N. Dilley, and L. Zhou, Phys. Rev. Lett. **95**, 106401 (2005).
[25] C. Stock, W. J. L. Buyers, Z. Yamani, C. L. Broholm, J.-H. Chung, Z. Tun, R. Liang, D. Bonn, W. N. Hardy, and R. J. Birgeneau, Phys. Rev. B**73**, 100504(R) (2006).
[26] D. A. Shulyatev, S. G. Karabashev, A. A. Arsenov, Y. M. Mukovskii, S. Zverkov, J. Crystal Growth, **237-239**, 810-814 (2002).
[27] J. M. deTeresa, M. R. Ibarra, P. A. Algarabel, C. Ritter, C. Marquina, J. Blasco,

J. Garcia, A. del Moral, and Z. Arnold, Nature (London) **386**, 256 (1997).

[28] Q. Huang, A. Santoro, J. W. Lynn, R. W. Erwin, J. A. Borchers, J. L. Peng, K. Ghosh, and R. L. Greene, Phys. Rev. B**58**, 2684 (1998).

[29] The energy resolution in these measurements is ~1 meV, which corresponds to a time scale ~1 ps.

[30] The higher-energy Jahn-Teller mode is also soft; see J. Zhang, P. Dai, J. A. Fernandez-Baca, E. W. Plummer, Y. Tomioka, and Y. Tokura, Phys. Rev. Lett. **86**, 3823 (2001).

[31] J. W. Lynn, J. Superconductivity: Incorporating Novel Magnetism **13**, 263 (2000).

[32] C. Sen, G. Alvarez, and E. Dagotto (preprint).

[33] R. Osborn, S. Rosenkranz, D. N. Argyriou, L. Vasiliu-Doloc, J. W. Lynn, S. K. Sinha, J. F. Mitchell, K. E. Gray, and S. D. Bader, Phys. Rev. Lett. **81**, 3964 (1998); S. Rosenkranz, R. Osborn, S. K. Sinha, K. E. Gray, J. F. Mitchell, L. Vasiliu-Doloc, J. W. Lynn, and D. Argyriou, J. Appl. Phys. **83**, 7348 (1998); S. Rosenkranz, R. Osborn, L. Vasiliu-Doloc, J. W. Lynn, S. K. Sinha, and J. F. Mitchell, Physica B **312**, 763 (2002); T. Chatterij, F. Demmel, G. Dhalenne, M.-A. Drouin, A. Revcolevschi, and R. Suryanarayanan, Phys. Rev. B**72**, 014439 (2005).

[34] A. S. Alexandrov, A. M. Bratkovsky, and V.V. Kabanov, Phys. Rev. Lett. **96**, 117003 (2006).

[35] J. C. Loudon and P. A. Midgley, Phys. Rev. Lett. **96**, 027214 (2006).

[36] B. I. Belevtsev, G. A. Zvyagina, K. R. Zhekov, I. G. Kolobov, E. Yu. Beliayev, A. S. Panfilov, N. N. Galtsov, A. I. Prokhvatilov, and J. Fink-Finowicki, Phys. Rev. B**74**, 054427 (2006).

[37] Recent specific heat data have been interpreted as a second order transition, uniquely contradicting other measurements that describe the transition at $T_C$ as first order. J. A. Souza, Y. Yu, J. J. Neumeier, H. Terashita and R. F. Jardim, Phys. Rev. Lett. **94**, 207209 (2005).

[38] J. Rodriguez-Carvajal, M. Hennion, F. Moussa, A. H. Moudden, L. Pinsard, and A. Revcolevschi, Phys. Rev. B**57**, R3189 (1998).

[39] Q. Huang, J. W. Lynn, R. W. Erwin, A. Santoro, D. C. Dender, V. N. Smolyaninova, K. Ghosh, and R. L. Greene, Phys. Rev. B**61**, 8895 (2000).

[40] K. Ghosh, C. J. Lobb, R. L. Greene, S. G. Karabashev, D. A. Shulyatev, A. A. Arsenov, and Y. Mukovskii, Phys. Rev. Lett. **81**, 4740 (1998).